\documentclass{aa}
\usepackage{graphicx}
\usepackage{txfonts}
\usepackage{longtable}
\usepackage{lscape}
\begin{document}
   \title{IPHAS extinction distances to Planetary Nebulae}


   \author{C. Giammanco\inst{1, }\inst{12}, S. E. Sale\inst{2}, R. L. M. Corradi\inst{1, }\inst{12}, M. J.  Barlow\inst{3},  K. Viironen\inst{1, }\inst{13, }\inst{14}, L. Sabin\inst{4}, M. Santander-Garc\'{\i}a\inst{5, }\inst{1, }\inst{12}, D. J. Frew\inst{6}, R. Greimel\inst{7}, B. Miszalski\inst{10 }, S. Phillipps\inst{8}, A. A. Zijlstra\inst{9}, A. Mampaso\inst{1, }\inst{12}, J. E. Drew\inst{2, }\inst{10}, Q. A. Parker\inst{6, }\inst{11}  \and R. Napiwotzki\inst{10}.}

   \institute{Instituto de Astrof\'{\i}sica de Canarias (IAC), C/ V\'{\i}a L\'actea s/n, 38200 La Laguna, Spain\\
              \email{corrado@iac.es}
        \and
                 Astrophysics Group, Imperial College London, Blackett Laboratory, Prince Consort Road, London SW7 2AZ, U.K.   \and
                  Department of Physics and Astronomy, University College London, Gower Street, London WC1E 6BT, UK \and
                 Instituto de Astronom\'{\i}a, Universidad Nacional Aut\'onoma de M\'exico, Apdo. Postal 877, 22800 Ensenada, B.C., M\'exico\and
                 Isaac Newton Group of Telescopes, Ap.\ de Correos 321, 38700 Sta. Cruz de la Palma, Spain \and
                 Department of Physics, Macquarie University, NSW 2109, Australia  \and
                 Institut f\"ur Physik, Karl-Franzens Universit\"at Graz, Universit\"atsplatz 5, 8010 Graz, Austria  \and
                 Astrophysics Group, Department of Physics, Bristol University, Tyndall Avenue, Bristol, BS8 1TL, U.K.  \and
                 Jodrell Bank Centre for Astrophysics, School of Physics and
                 Astronomy, University of Manchester, Oxford Road, M13 9PL, Manchester, U.K.  \and
                Centre for Astrophysics Research, STRI, University of Hertfordshire, College Lane Campus, Hatfield, AL10 9AB, U.K.  \and
                Anglo-Australian Observatory, PO Box 296, Epping, NSW 1710, Australia \and
                Departamento de Astrof\'\i sica, Universidad de La Laguna, E-38205 La Laguna, Tenerife, Spain \and
                Centro Astron\'omico Hispano Alem\'an, Calar Alto, C/Jes\'us Durb\'an Rem\'on 2-2, E-04004 Almeria, Spain \and
                Centro de Estudios de F\'{\i}sica del Cosmos de Arag\'on (CEFCA), C/General Pizarro 1-1, E-44001 Teruel, Spain
 }

   \date{Received September 15, 1996; accepted March 16, 1997}

 
  \abstract
  {}
  {The determination of reliable distances to
  Planetary Nebulae (PNe) is one of the major limitations in the study of this
  class of objects in the Galaxy. The availability of new photometric surveys
  such as IPHAS covering large portions of the sky gives us the opportunity to
  apply the ``extinction method" to determine distances of a large number of
  objects.}
   {The technique is applied to a  sample of 137 PNe located between -5 and 5 degrees in Galactic latitude, 
and between 29.52 and 215.49 degrees in longitude. The characteristics of the distance-extinction method and the main sources of errors are carefully discussed.}
   {The data on the extinction of the PNe
   available in the literature, complemented by new observations,
   allow us to determine extinction distances for 70 PNe. 
    A comparison with statistical distance scales from different
     authors is presented.}\\



   {}

 \authorrunning{Distances of Planetary Nebulae}
   \maketitle
%

\section{Introduction}

Distances to Galactic planetary nebulae (PNe) present a severe
and longstanding problem. Obtaining an accurate distance scale for PNe
will allow us to compute the total number of PNe in the Galaxy, which
has important implications for the Galactic ultraviolet radiation
field, the total processed mass returned to the interstellar medium,
and more generally to our understanding of the chemical evolution of
the Galaxy.

A method that is a priori independent of assumptions about the
physical or geometrical properties of the nebulae is the extinction
method.  Assuming that the interstellar extinction to a certain nebula
can be determined, if one is able to build up the extinction-distance
relation using field stars around the line of sight to the nebula, the
relation can be used to infer the distance to the PN.

The application of this method is not new. Lutz (1973) measured the
distance to 6 PNe using some 10 field stars per object. Later, the
method was applied by Acker (1978) who provided reliable distance
 values for 11 PNe and a rough estimation for 34 other ones, and more
recently by Gathier et al. (1986) who measured the distance to 12 PNe
using some 50 stars per PN.  In the last paper, a comprehensive
discussion about different aspects of this method can be found. The
number of stars for determining distances was increased by Pollacco \&
Ramsay (1992) who, using a colour analysis of the field stars, were
able to reach an accurate spectral classification for stars later than
F5 type. As late-type stars constitute the most numerous objects in
all galactic-plane directions, they allow an extensive application of
this method to determine distances to PNe in the Galaxy.

The availability of the IPHAS H$\alpha$ survey of the Northern
Galactic Plane (Drew et al. 2005) and its coming extension to the
South (VPHAS+) opens new doors for the application of the method.
 IPHAS allows us to determine extinction-distance curves using a
large number of field stars, typically several hundred  in
areas as small as $10\arcmin \times 10\arcmin$ around each line of
sight. The technique is presented and discussed by Sale et
al. (2009). In this paper, we will present its application to Galactic
PNe. We determine the distance to 70 PNe included in the
ESO/Strasbourg catalogue (Acker et al. 1992), and compare the
distances obtained with those obtained by other authors using
different methods.  This shows how IPHAS and its successor
surveys potentially provide a new and powerful tool to obtain
distances to a large number of Galactic PNe.
 
\section{The data}
\label{data}

IPHAS is a wide-field, CCD, H$\alpha$ survey of the Northern Galactic Plane (Drew et
al., 2005), carried out at the 2.5m Isaac Newton Telescope on La Palma,
Spain. Imaging is performed also in the r$\arcmin$ and i$\arcmin$
bands down to r$\arcmin \sim 20$ (10$\sigma$). The high quality
photometry and characteristics of the survey permit the spectral
classification of main-sequence stars to be determined, based on the H$\alpha$ line
strength. 
The availability of reliably calibrated (r$\arcmin$ - H$_{\alpha}$) colours permit the
spectral classification of most stars, while the (r$\arcmin$- i$\arcmin$) colour
provides their extinctions, allowing distances to be estimated from the
r$\arcmin$ measurements. The limiting magnitude in
r$\arcmin$ permits extinctions to be measured to distances of up $\sim
10$ kpc.

IPHAS covers the Galactic latitude range between -5 and 5 deg, and longitude
range between 29.52 and 215.49. In this region of the
sky there are 190 known PNe according to the ESO/Strasbourg catalogue
(Acker et al., 1992).  We were able to extract a reddening for 137 of them from
different sources; for 27 PNe we present the first determinations.

\subsection{PN extinctions extracted  from the literature}
\label{av}

The ESO/Strasbourg catalog provides the ${\rm H_{\alpha}}/{\rm
  H_{\beta}}$ line ratio for most of the PNe considered here. 
  These data were measured from spectra obtained using two different telescopes:
  the 1.52m ESO telescope for the Southern nebulae, and the 1.93m OHP
  telescope for the Northern ones. Further details about the
  instruments (photographic plate or CCD) and spectra can be found
  in Acker et al. (1992, SECGPN), Acker \& Stenholm (1987), and Stenholm \& Acker (1987).  
Later, Tylenda et al. (1992) used these SECGPN line intensities to
obtain the value of the extinction constant roughly determined for
about 900 PNe.  In this paper the visual extinction was obtained from
  the line ratios by applying the formulae:
\begin{equation}
c_{\beta}=2.84 \cdot \log  \left( \frac{(\rm {\rm H_{\alpha}}/{\rm H_{\beta}})}{ 2.86} \right)
\label{f1}
\end{equation}
\begin{equation}
{\rm A_v}=2.15 \cdot c_{\beta}
\label{f2}
\end{equation}
which come from Fitzpatrick (2004). Here we assumed R=3.1,
in order to be consistent with the method by which the 
IPHAS extinction curves are built (see section \ref{stuart}).

Other compilations from which $ c_{\beta}$ was retrieved are Cahn et
al. (1992) and in few cases Stasi\'nska et al. (1992). The latter paper
compared the optical and radio extinction determinations, using 
the same ``SECGPN'' line fluxes as Tylenda et al. (1992), and adding some other measurements, 
for a total of 130 PNe.
For individual nebulae, these listings where supplemented with data from
  other papers, where available.  In all cases, when in a paper the assumed
  reddening law is reported, we normalize the published extinction using
  Eqs. (\ref{f1}) and (\ref{f2}). Specifically, this is the case for data in
  Cahn et al. (1992) and Stasi\'nska et al. (1992).  For most of the nebulae
the only available extinction coefficient comes from the Strasbourg catalog,
while for a subset there are multiple determinations. For a small number of
nebulae there is a very extensive literature; and in those cases we selected what
we believe is the most accurate determination of $c_{\beta}$ after a careful analysis
of bibliographic sources. In addition, for some nebulae we also present new
optical $c_{\beta}$ determinations. 

A general rule in case of multiple determinations and in the absence of other
information (the quality of one measurement with respect to another) is to
choose the smallest $c_{\beta}$.  The main reason is that slit spectra taken
at non-negligible zenith distances are often affected by differential
atmospheric dispersion which removes blue light leading to an overestimate of
the $c_{\beta}$'s.  Extinction measurements obtained via radio data are free
from this effect, but these give extinctions
systematically lower than those determined from the Balmer decrement
(Stasi\'nska et al., 1992, Ruffle et al. 2004).
In order to deal with a homogeneous set
of measurements, we will only consider in this paper $c_{\beta}$
obtained via optical measurements.  

\begin{figure}
\resizebox{\hsize}{!}{\includegraphics{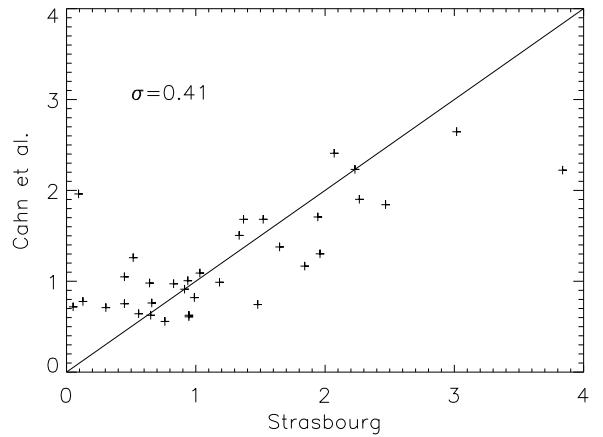}}
\caption{ $c_{\beta}$ measurements by
  by Cahn et al. (1992) plotted against data in Acker et al. (1992). The dispersion is around 0.4 in $c_{\beta}$.}
\label{fig1}
\end{figure}

Finally, in order to illustrate the differences between extinction
determinations by different authors, we show in Fig. \ref{fig1} the
$c_{\beta}$ values obtained by Cahn et al. (1992) versus those
computed from fluxes in the ESO/Strasbourg catalog. The associated
scatter is $\sigma=0.41$, which shows how critical the choice of the
extinction values can be.  We note that
data from the ESO/Strasbourg catalog are not all of the same
  quality, those for the Northern nebulae ($ 56< l< 184$) observed at
  the OHP being generally of lower quality than those of (mainly
  Southern) nebulae observed at ESO.

\subsection{New extinction determinations}

26 PNe considered in this paper (see Tabs. 1 and 3) were observed at
the 2.5m Isaac Newton Telescope at the Observatorio del Roque de los
Muchachos, from August 23 to 30, 1997.  The IDS spectrograph
was used, together with its R300V grating, which gives a reciprocal
dispersion of 3.31~\AA\ per pixel of the 1kx1k Tek3 detector, and a
spectral coverage from 3700 to 6900~\AA.  The slit width was
1.5~arcsec projected on the sky, providing a spectral resolution of
6.7~\AA. Total exposure times on each nebula varied from a few seconds
for the brightest target (NGC 7027) up to one hour for the faintest
ones; exposures were split in order to have both the bright and faint
nebular lines well exposed without saturation.  Several
spectrophotometric standards were observed during each night for
relative flux calibration. Data reduction was performed using the
package {\it onedspec} in IRAF.

An additional two PNe (PC 20 and Sa 3-151) were observed with the Dual
Beam Spectrograph (DBS, Rodgers et a.1988) on May 10 and 14 2008 by
B. Miszalski at the Australian National University 2.3-m
telescope. The exposure times were 300 s and 150 s, respectively to
achieve a SNR of about 20 in the peak intensity of ${\rm
  H_{\beta}}$. The 1200B and 1200R gratings were used with a slit
width of 2\arcsec (positioned at $PA=270$ deg) to give wavelength
coverage windows of 4030-5050~{\AA} and 6245-7250~{\AA} at a
resolution of $\sim$1.6~{\AA} (FWHM). Data reduction was performed
using IRAF and a number of flux standards were observed each night to
derive the spectrophotometric response across the separate blue and
red spectrograph arms.

\section{IPHAS extinction-reddening curves for the PNe}
\label{stuart}

Extinction-distance relationships for the lines of sight toward the
sample of 136 Galactic PNe in the IPHAS area have been computed
with the algorithm MEAD, as described by Sale et al. (2009). MEAD exploits a feature of the
IPHAS colour-colour plane, whereby it is possible to simultaneously
accurately determine the spectral type of a star and its reddening,
avoiding the serious degeneracies that
exist in other filter systems and with only a slight dependence on a Galactic model.

Typically, several thousand A to K4 stars in a box of sides 10$\arcmin$,
centred on the PNe, have been employed in the computation of each
relationship. Distances to each star are obtained from their 
apparent magnitude, the MEAD extinction , and the absolute
magnitude of the star derived from the MEAD spectral
type. The photometric errors are propagated to give
an error on the estimated distance to each star. The stars were then
binned by distance, with each bin being at least 100~pc deep, containing
at least 8 stars and having a total signal to noise ratio in the bin of at
least 130. In a small departure from the method described in Sale et al.
(2009), the prior probabilities of each luminosity class are determined on
a sightline by sightline basis. MEAD returns monochromatic extinctions,
these have then been converted to ${\rm A_v}$ using the R=3.1 reddening
law of Fitzpatrick (1999).


\section{The inversion of Distance-Extinction curves}
\label{det}

\begin{figure}
\resizebox{\hsize}{!}{\includegraphics{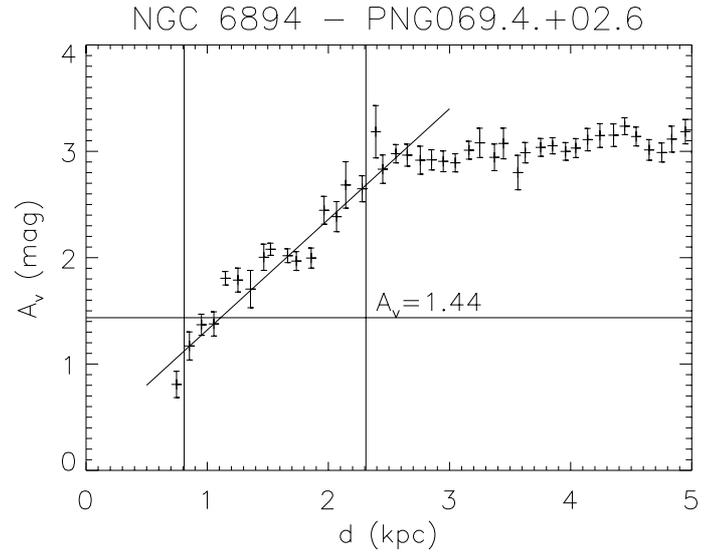}}
\caption{Extinction law within $10\arcmin \times 10\arcmin$ around the line of
  sight toward the NGC 6894.  The measured extinction of the
  nebula is also indicated by the horizontal line, and the two vertical lines limit the
  interval in which the law is linear and which  we have used to find
  the corresponding distance to the PN.}
\label{fig4}
\end{figure}

The extinction-distance curves  obtained are in general linear
for a significant range of distances, until they generally flatten out
and reach an asymptotic value once the line of sight leaves the dust layer 
within the Galactic Plane (Fig. \ref{fig4}). Even  in the cases where
they show a more complex shape, it is generally possible to select a
linear region around the point in which we are interested, e.g. the
${\rm A}_{\rm v}$ measured for the nebula.

We assume that the measured ${\rm A}_{\rm v, i}$ are distributed
around a true value ${\rm A}_{\rm v, i}^t=a+bd_{\rm i}$ following a
Gaussian law whose dispersion is $\sigma_{\rm A}$.  If so, a least square fit
to the selected part of the curve allows us to determine all the
needed parameters. Specifically, around a point, for each $d$, the
probability-distribution to measure a given A$_{\rm v}$ is P(${\rm
  A_{\rm v}}|d$).  The next step is to invert the probability distribution,
i.e. to obtain the probability to find a distance $d$ for a given
A$_{\rm v, i}$: P($d|{\rm A_{v, i}}$).

We note that:
\begin{equation}
{\rm P}(d|{\rm A}_{\rm v, i})=\frac{{\rm P}({\rm A}_{\rm v, i}|d)}{\int {\rm P}({\rm A}_{\rm v}|d) {\rm d{\rm A}_{\rm v}}} 
\end{equation}
and using the previous assumptions of linearity and Gaussian distribution for P(${\rm A_v}|d$), 
the new distribution for P($d|{\rm A_v}$) is itself Gaussian. 
We call the central value of this distribution $d_{\rm A}$ and $\sigma_d$ the dispersion. 
It follows that
\begin{equation}
  d_{\rm A}=(A_{\rm v, i}-a)/b,
 \end{equation}
and
\begin{equation}
 {\sigma}_d={\sigma}_{\rm A}/b.
 \end{equation}

In Fig. \ref{fig4}, we show as an example the well defined
extinction curve along the line of sight towards NGC 6894.  For this
nebula, the measured extinction value is $c_{\beta}=0.67 $, or ${\rm
  A_v}=1.44 $ (data from Ciardullo et al., 1999).  The limits for the selected linear range are 800 and
2300 pc (sensitivity range of the method). The estimated distance is  $d=1000$ pc and the
formal standard deviation is ${\sigma}_d=150$ pc. We stress that the
formal ${\sigma}$ is only part of the total uncertainty, whose major
contribution is due to the error in $c_{\beta}$ measurement; 
other uncertainties we will consider in the following section.
\begin{figure}
\resizebox{\hsize}{!}{\includegraphics{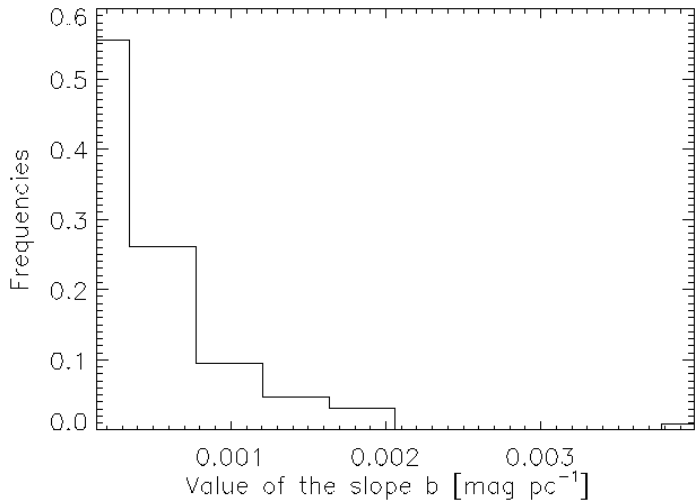}}
      \caption{Distribution of the values of the slope $b$ mag/pc obtained from a linear fit of the
        interstellar extinction curves. We analyzed more than
        100 lines of sight corresponding to the know PNe in the IPHAS
        survey. Almost 80\% of the values are distributed between
        1.3$\cdot 10^{-4}$ and 8$\cdot 10^{-4}$ }
\label{figb}
\end{figure}

In order to simplify the discussion in
the next section about the uncertainty introduced by the inaccuracies in
${\rm A_v}$, we will fix  representative values the parameters $b$
and  ${\sigma}_d$. This is done by analysing almost
two hundred lines of sight. We found a representative mean value for
the parameter $b$ of 6.53$\cdot 10^{-4}$ mag/pc (Fig.\ref{figb}), and a
typical error in the fitting process of around $10\%$. A test distance of
2500 pc is also assumed. This is in the middle of the range over which
the extinction-distance method is sensitive 
for the low Galactic latitude coverage that characterizes the IPHAS survey.
This range was calculated by taking the mean of the lower and upper
limits of the linear interval corresponding to each curve. For the adopted
distance, the typical error for the fitting process is around 250 pc. If so,
we can express the relative uncertainty for the distance as:

\begin{equation}
E=\frac{\sqrt{250^2 + (\sigma_{\rm Av}/6.53~10^{-4})^2}}{2500}
\end{equation}
\begin{equation}
E_{\rm Av}=\frac{\sigma_{\rm Av}/6.53~10^{-4}}{2500}
\end{equation}
where $E$ is the total uncertainty and $E_{\rm Av}$ is the
contribution of ${\rm A_v}$.  
From the previous formulae for $\sigma_{\rm Av}$=0.16
($\sigma_{c\beta}$=0.077) we obtain $E_{\rm Av}\sim$10\% and
$E\sim$14\%.  Finally we note that the previous formulae are
exact in the case of purely statistical errors, otherwise they represent a lower
limit to the true uncertainty. 

\section{On the errors}
\label{errori}

In section \ref{data} we saw how large the difference can be
between the $c_{\beta}$ values determined for the same nebula by
different authors. The origin of these differences could be systematic
errors due to the instruments or in the reduction
processes. Unfortunately it is impossible to quantify {\it a priori}
this effect. However, in the determination of the interstellar
reddening and hence of the extinction-distance there are other sources
of uncertainty that we discuss below.

\subsection{Sources of error relevant for all objects}
\label{eer}
 
In order to determine the visual extinction to a PN, it is common to
use the ${\rm H}_{\alpha}/{\rm H}_{\beta}$ ratio. However, often the
literature data quote only the extinction coefficient $c_{\beta}$. In
some cases, the adopted extinction law is also quoted, in such a way
that we can convert the c$_{\beta}$'s to a common scale, that we adopt
to be the law given in Fitzpatrick (2004), for R=3.1, for a
theoretical ratio ${\rm H_{\alpha}}/{\rm H_{\beta}}$=2.86.

The first uncertainty that we consider is that associated with the
theoretical ${\rm H_{\alpha}}/{\rm H_{\beta}}$ ratio as a function of
the electron temperature, for case B. Using Osterbrock (2005) for
this ratio we find that if the electron temperature varies 
between 8000 and 20000 K, the largest error that we would make by
adopting a ratio of 2.86 is 0.04 dex in $c_{\beta}$, or 0.09 mag in
visual extinction. This implies an uncertainty of up to 11\% in
the distance determination.

The second error source that we consider is the adopted extinction
law. Several authors use extinction laws different from that adopted in
this paper. 
For example, Stasi\'nska et al. (1992) used the Seaton
(1979) law assuming R=3.2, Cahn et al. (1992) used the Whitford (1958) law
(R=3.2), while the Howarth (1983, R=3.1) law was used by other authors. The
maximum difference between our adopted law occurs when the Cardelli et al. (1989)
 law is used. We conservatively adopt this difference when the
extinction law is not quoted in a paper from which the $c_{\beta}$ for
a specific PN is adopted.  This results in a difference of 15\% in
$c_{\beta}$, or $\sigma_{\rm Av}$= 0.3 (for A$_{\rm v}$=2), and a
contribution to the error on the distance of about 20\%.

Another common hypothesis is that the characteristics of the
interstellar dust are constant for all lines of sight, and correspond
to $R=3.1$, the mean value of the Milk Way. Fitzpatrick (1999) and
Fitzpatrick \& Massa (2007) state that for optical wavelengths a
single value of R is a good approximation, but $R$ can vary for
different lines of sight. Several authors report $R\sim2$ through the
galactic bulge (Byun 1996, Udalski 2003, Ruffle et al. 2005) and towards the halo
(Larson and Whittet, 2005). Stasinska et al. (1992), based on the
study of a sample of PNe, suggest that for the Galactic Plane a more
appropriate value is $R=2.5$.  The parameter $R$ could be instead
significantly larger if the line of sight crosses a dense dust cloud
(due to the generally larger size of the grains).  However, it
  would be unlikely to be able to detect the blue part of the spectrum
  of PNe (and thus their ${\rm H}_{\beta}$ flux) along such highly
  extinguished directions. In general, the $R$ value for a particular
nebula can in principle be determined if the number of Balmer lines
measured with precision is large enough.

Therefore, we cannot exclude variations of $R$ for our IPHAS sample. Having
adopted $R=3.1$ as a common value, we estimate the error that we make if the
true extinction ratio were 2.1 or 5.5 (maximum and minimum values from
Fitzpatrick, 1999 and Fitzpatrick \& Massa 2007). For the case $R=2.1$, the
Fitzpatrick law gives ${\rm A_v}=4.8 \cdot \log({\rm H_{\alpha}}/{\rm
  H_{\beta}}/2.86)$ while we use ${\rm A_v}=6.1 \cdot \log({\rm H_{\alpha}}
/{\rm H_{\beta}} /2.86)$. This results in an overestimate of ${\rm A_v}$ by
some 21\% , or $\delta_{\rm Av}$= 0.4 (for A$_{\rm v}$=2). However, in this
case the corresponding distance extinction curve is also overestimated by
about 15\% (section \ref{stuart}). This means that when determining distances
these errors partially compensate and the corresponding overestimate of ${\rm
  A_v}$ is of about 6\% , or about 12\% for a distance of 2500 pc. For $R=5.5$,
we underestimate $A_v$ by some 33\% , but the corresponding extinction curve is
also underestimated by 25\% , so that a final compensated error is 8\% in
${\rm A_v}$, or $\sim$14\% for the distance.

\subsection{Error coming from the object's properties}

An additional uncertainty resides in the hypothesis that the
interstellar medium is responsible for all the reddening measured for
a PN from the Balmer decrement, in other words, that the PNe are free
of internal or circumnebular absorbing dust (for a review see Barlow, 1983). This
hypothesis is in general supported by the study of K\"oppen (1977) 
which, assuming that in a planetary nebula the dust and gas components are well
mixed, finds that the nebular dust optical depth is very small
(${\tau}<0.05$), for a dust/gas ratio around $10^{-3}$.  However this
could be too simple a picture, and in this original work there were also some 
PNe (6 out of a total of 21) with evidence for
associated extinction. The effect is probably negligible for standard,
elliptical PNe (Barlow, 1983), but could be more significant for
particular nebular morphologies, such as that of bipolar PNe 
from massive progenitors (e.g. Corradi \& Schwarz 1995) which can have
massive neutral equatorial envelopes. There is some evidence in
the literature that this might be the case.

The most studied example of the latter type is NGC 7027. Osterbrock
(1974) estimated a maximum internal absoption of 0.6 mag. Woodward et
al. (1992) suggested that the obscuring dust lies in a shell or disc
external to the ionized gas, and recently Bieging et al. (2008) found
variations of $c_{\beta}$ from 0.8 to 2.4. A variation of A$_{\rm v}$
has been measured also for NGC 650-1 between the SW and NE limits of
the central emission bar by Ramos--Larios et al. (2008). Other type-I
bipolar nebulae, such as Sh 2-71, K 3-94, K 4-55 and M 1-75,
show some evidence for (generaly modest) extinction variations through
their structures (Bohigas 1994, 2001). Extreme variations are reached for
NGC 6302 and NGC 6537 (Matsuura et al. 2005a, 2005b).

There are also a few indications  of associated
extinction for non-bipolar morphologies\footnote{Note the
    possible morphological misclassification of a fraction of PNe due
    to orientation effects: see Manchado (2004)}. In this context an
interesting object is NGC 6741, for which Sabbadin et al. (2005) find
a circumnebular neutral halo generated during a recombination phase
following the fading of the central star. This halo was estimated to
be responsible for 10-20\% of the measured total $c_{\beta}$. The
Helix nebula (NGC 7293) seems to be a very complex object where dust,
ionized and molecular gas cohabit (Speck et al. 2002). Dense knots in
a PN could also evolve in absorbing filaments, as reported by O'Dell
et al. (2003) for the bipolar PN IC 4406.

For NGC 6781, Mavromatakis et al. (2001) provide a 2D Balmer decrement
map, finding variations of ${\rm H_{\alpha}}/{\rm H_{\beta}}$ ranging
from 5.6 to 7.   Though smaller, similar structures are also
visible in a similar map obtained for Menzel 1 by Monteiro et
al. (2005).

Summarizing, in some cases  care has to be taken when
determining the interstellar extinction, especially for particular
geometries (bipolar). However, due to the small number of studies it
is difficult to estimate the percentage of objects belonging to these
categories whose interstellar extinction could be wrongly estimated. In any case the
number of bipolar PNe is 15\% of the total number of planetary nebulae
(Corradi \& Schwarz, 1995), and from
this sub-sample significant unrecognized errors are probable only for distant, 
spatially unresolved objects. We can conclude that only a small fraction
of our total sample will be affected by this kind of uncertainty in
the estimate. \\

 Concluding, at present the major contribution to the uncertainty
  on the distances is due to the $\rm A_v$'s error measurements. For
  two nebulae, NGC 6842 and NGC 7048, our own measurements allows us to estimate these errors. 
Combining it to the
  statistical fitting error leads to a final uncertainty on the
  distance of about 35\%. 


\section{Distance determinations for the NGC sample}
\begin{table*}[!t]
\caption{Distances inferred for the NGC sample. Columns labelled as
  ``Stras", ``Cahn'' and ``Stas'' are referred to the Strasbourg
  catalog, Cahn et al. (1992) and Stasi\'nska et al. (1992), as well
  as ``dC'' and ``dVst'' are referred to Cahn et al. (1992) and van de
  Steene \& Zijlstra (1994), respectively, while ``dMa'' is referred
  to Maciel (1984). ``Other" references are given in the text. 
    When significantly different extinction estimates are available, 
    the  values reported for the distances are
    obtained using the absorption values in bold face. Errors 
    are the statistical ones coming from the linear fit on
    the distance extinction curve, except for NGC$~$6842 and NGC
    7048, where we have also included the errors on the
    extinction measurement. }
\label{tav1}  \centering %
\begin{tabular} { l c | c c c c c| c c c r} 
  PNG & Name &   \multicolumn{5}{c}{$\rm A_v$} &  \multicolumn{4}{c}{distance [pc] } \\
           &	        &    Stras.  &    Cahn &  Stas.  & this paper & Others &  dC &   dVst &  dMa & this paper      \\
\hline
\hline                 
033.8-02.6  &  NGC 6741 &      1.96    &     1.96  &     1.93  &                                   &\bf 1.73 &    2047 &  2260 &     1700   &  2680 $\pm$ 110 \\
041.8-02.9  &  NGC 6781 &                 &     2.09  &     1.96  &  2.16 $\pm$ 0.11    &     1.78 &     699  &    840 &   900    &  $<$ 1000\\
045.7-04.5  &   NGC 6804 &                &     1.85  &              &  1.85 $\pm$ 0.11           &\bf 1.58 &   1709  &  1660 &      1600       &   $<$   800  \\ 
046.4-04.1  &   NGC 6803 &      0.96   &     1.62  &     1.60  &                                   &\bf 1.14&    2987  &  3190 &      2500        &    950 $\pm$ 130 \\
060.8-03.6  &   NGC 6853&                 &      0.36 &              &                                   &    0.55 &    262   &  400   &        400   &  $< $ 800 \\       
065.9+00.5 &   NGC 6842 &      0.96   &     2.04  &              &    {\bf 2.08 $\pm$ 0.15}       &              &    1366  &            &   1700 &     2700 $\pm$ 950 $\dagger$ \\  
069.4-02.6  &   NGC 6894 &                &     1.80  &              &  1.85 $\pm$ 0.11          &\bf 1.44 &   1653  &  2000 &    1500 &      1000 $\pm$ 100  \\
074.5+02.1 &   NGC 6881&   3.27       &      3.62 &              &                                   &     4.12 &    2473  & 3950   &  1700  & n.d. \\        
084.9-03.4  &   NGC 7027&                 &     2.53  &              &   2.97 $\pm$ 0.46          &              &     273 &    630 &     700 &  n.d.\\
088.7-01.6  &   NGC 7048 &  0.65       &     1.53 &               &   {\bf 1.20 $\pm$ 0.13}   & 1.27  &     1598 &  2220 &  1200 &      1000--2000 $\dagger$ \\
089.0+00.3  &  NGC 7026 &  1.40       &     1.34  &              &   2.10 $\pm$ 0.11          &  1.69&    1902 &  1940 &     900 &   $<$   1500 \\ 
107.8+02.3  &  NGC 7354&   2.94       &     3.62  &              &   3.99 $\pm$ 0.08          &\bf 3.23 &    1271 &  1230 &    800  &  1000 $\pm$ 150  \\
\hline
\end{tabular} 
\end{table*} 
\begin{figure*}
\centering
\resizebox{\hsize}{!}{\includegraphics{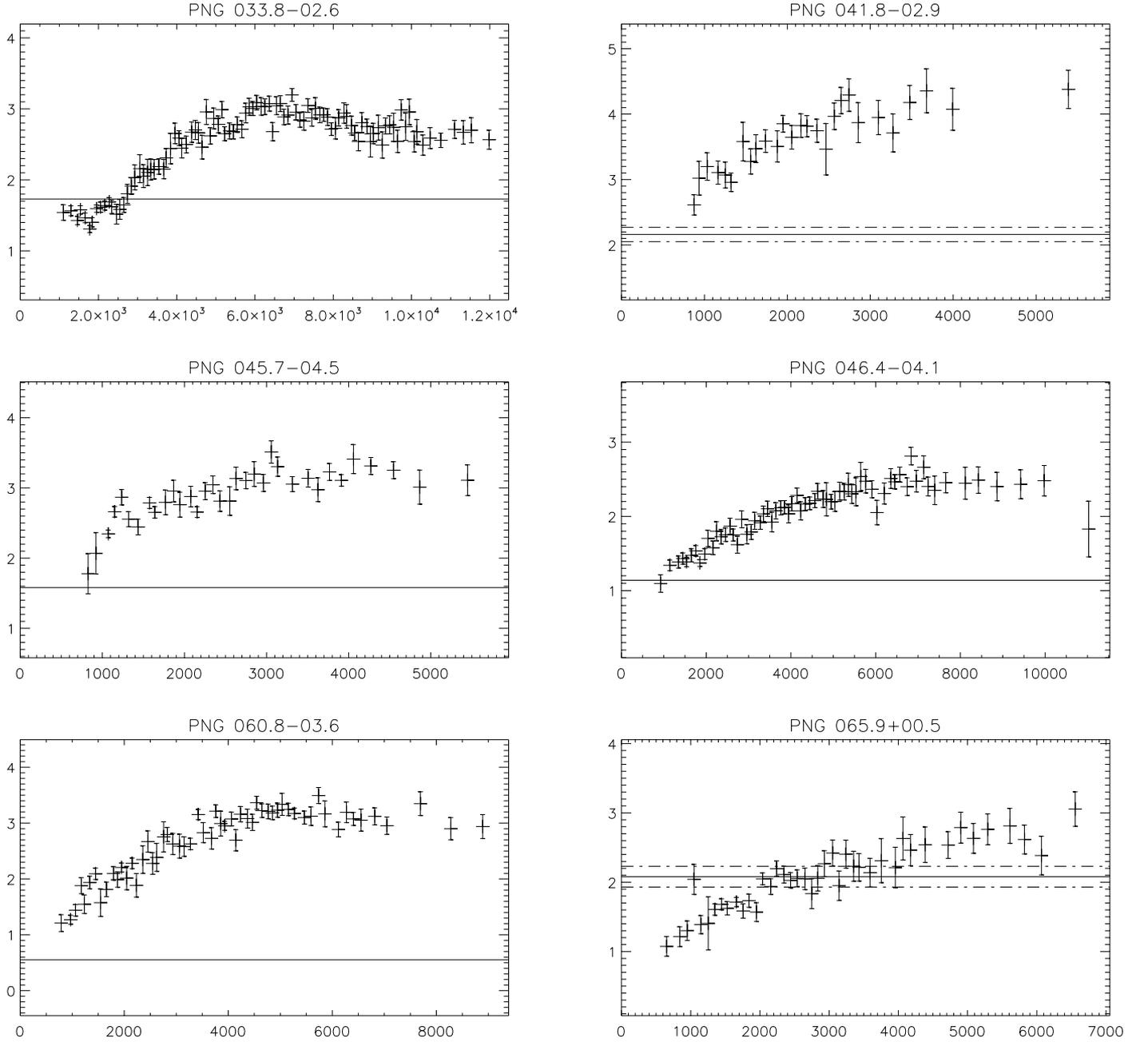}}
\caption{In the figure we present the extinction distance curves for the first 6 nebulae reported in Table \ref{tav1}. The abscissae  are in pc, the ordinates, in visual magnitudes and the horizontal lines represent the visual extinction corresponding to each nebula, as discussed in the text.  The short dash-long dash lines represent the 1-$\sigma$ error associated to our A$_{\rm v}$ measurement presented in Table \ref{tav1}.}
\label{lista1}
\end{figure*}
\begin{figure*}
\centering
\resizebox{\hsize}{!}{\includegraphics{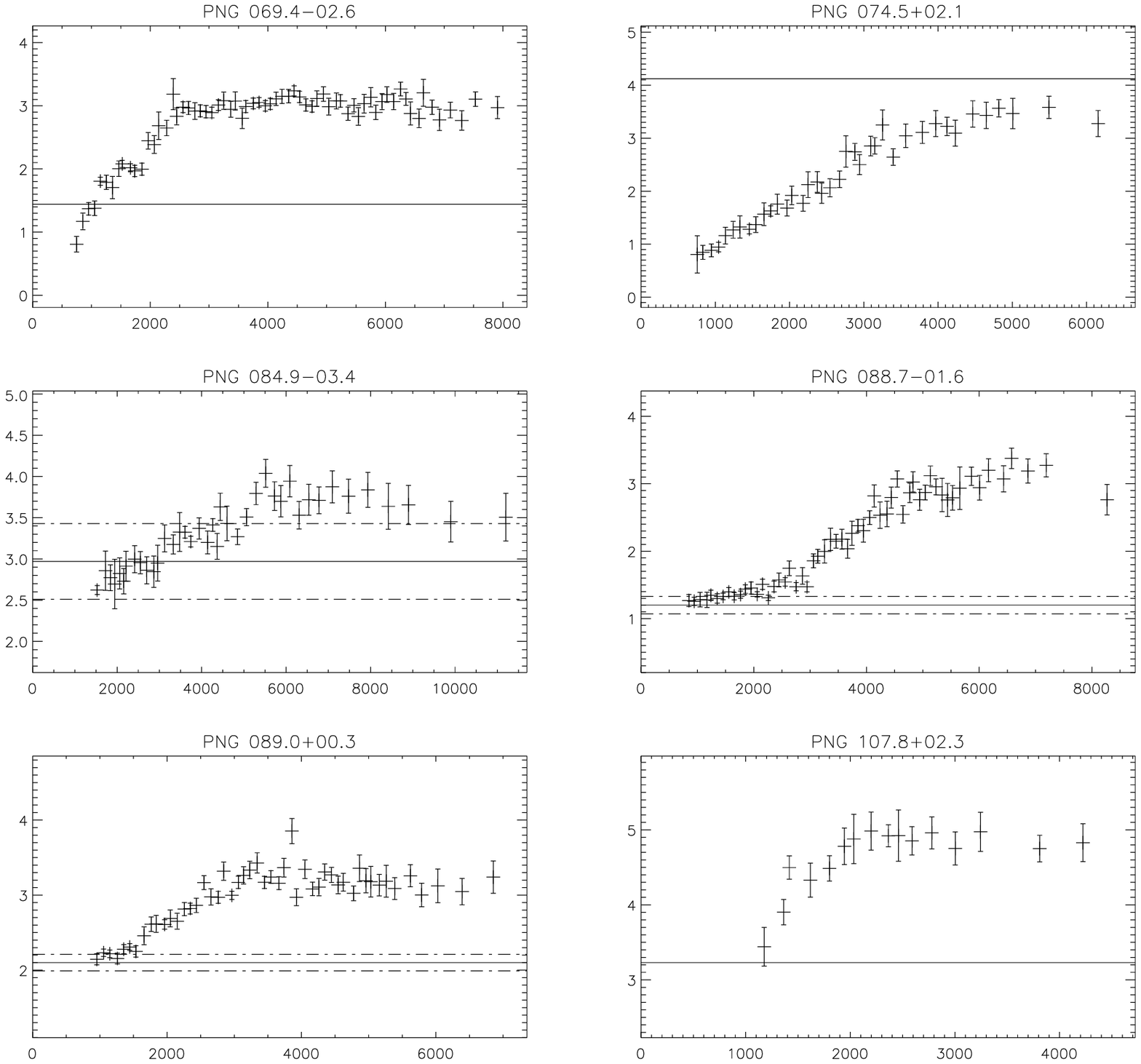}}
\caption{Extinction-Distance  curves for the second group of six nebulae presented in Table \ref{tav1}.  For all graphics the abscissae  are in pc, the ordinates, in visual magnitudes. The horizontal lines represent the visual extinctions corresponding to the nebulae considered in the text.  The short dash-long dash  lines represent the 1-$\sigma$ error associated to our A$_{\rm v}$ measurement presented in Table \ref{tav1}.} 
\label{lista2}
\end{figure*}

In order to minimize the uncertainties we will infer the distances for a
sample of very well known PNe in the IPHAS area, shown in Table~\ref{tav1}.

The first PN is {\bf NGC 6741 (PNG 033.8-02.6, Fig. \ref{lista1})}
which, as discussed in the previous section, was found to have
non-negligible internal extinction by Sabbadin et al. (2005). We use
the interstellar value given in the same paper to determine its
distance.  The reported error on the distance comes from the
  statistical $\sigma_d$. This also applies to the other PNe in this section, except 
for NGC 6842 and NGC 7048 as discussed later.

For {\bf NGC 6781 (PNG 041.8-02.9, Fig. \ref{lista1})}, adopting
all the reported values of $\rm A_{v}$, from the minimum value
inferred by Mavromatakis et al. (2001) to the maximum value obtained
in this paper, we get a distance smaller than 1000 pc. We can not be
more precise since the curve is not determined within this range.
However, the result is in line with that of Schwarz and Monteiro
(2006), who found a distance of 750 pc using 3D photoionization
modeling.

For {\bf NGC 6804 (PNG 045.7-04.5, Fig. \ref{lista1})}, we use the
determination of $\rm A_v$ obtained by Ciardullo et al. (1999), using Hubble
images. The graph has been transfered to the Fitzpatrick law ($R=3.1$), which
is our adopted law for the interstellar extinction. In this case, the inferred
distance is $< 800$ pc, while from our measurement of $c_\beta$ and the
  corresponding error we estimate $850 \pm 70$ pc.  This is a factor of 2
smaller than the statistical distances reported by Cahn et al. (1992) and van
de Steene \& Zijlstra (1994).  The error on the statistical scales is
typically 30\%\ (1$\sigma$), so that these values do not exlude the much
shorter (and with a smaller uncertainty) distance found here.  In favour of
the smaller distance, we also note that the risk of the extinction method is a
tendency to over-estimate distances. Moreover, this is a bipolar nebula for
which it may will be possible that the measured extinction is significantly
bigger than the interstellar component due to absorbing structure associated
with the PN.

In our method, the only circumstance in which we can  underestimate the distance is
if the parameter $R > 3.1$.
This can occur when the line of sight intersects a dense molecular cloud (a rare event).
The map of dense molecular clouds in the Galaxy given by Hartmann \& Thaddeus
(2001)  does not show  any clouds at these coordinates.
So we recommend the use of our
estimated range $<800$ pc, which is also near the estimate of 870 pc by
Frew (private communication, Frew et al.  2010 in preparation), based on a relationship between size and
surface brightness for PNe.

For {\bf NGC 6803 (PNG 046.4-04.1, Fig. \ref{lista1})}, using the extinction
value determined by Peimbert \& Torres-Peimbert (1987) ($c_{\beta}$=0.53, $\rm
A_v=1.14$), we infer a distance of 950 pc, again less than half the previously
reported estimates. Even using the much higher values of $c_{\beta}$ presented
by Cahn et al. (1992) we obtain a smaller distances ($2100 \pm 300$ pc) than
those previously reported.

For the nearby nebula {\bf NGC 6853 (PNG 060.8-03.6, Fig. \ref{lista1})}, there exists a parallax measurement
(Benedict et al., 2003) yielding  a distance of $417^{+49}_{-65}$
pc and a visual extinction coefficient tabulated in column labelled ``Others" of
Tab. \ref{tav1}. In this case the nebula is outside of the sensitivity
range of our method and we can give only an upper limit to the distance.

For {\bf NGC 6842 (PNG 065.9+00.5, Fig. \ref{lista1})}, there is
little information in the literature. From our own measurement
obtained at the 2.5m Isaac Newton Telescope on La Palma, we 
determine an extinction measurement that is clearly higher than that
inferred by adopting line fluxes from the Strasbourg catalog.
Observations imply a distance of 2700$\pm 950$ pc.  The error on
  the distance has been calculated taking into account the fitting
  $\sigma_d$= 749 pc, the error on the extinction measurement 0.15,
  and the slope of the curve, $b=2.6\cdot 10^{-4}$ pc/mag.

{\bf NGC 6881 (PNG 074.5+02.1, Fig. \ref{lista2})} is a quadrupolar young
nebula (Guerrero \& Manchado, 1998) suggested by Sabbadin et al. (2005) to
have been produced by a massive progenitor and presently in a similar
recombination phase to NGC 6741. If so, it is possible that $c_{\beta}$
overestimates the interstellar extinction and thus the distance using the
extinction method, which would then be greater than $4000$ pc using the very
high absorption reported in column ``Others'' (Kaler \& Kwitter, 1987), as
well as the value obtained from the Strasbourg catalog. For this reason we do
not quote a distance in Tab. \ref {tav1}.
  
{\bf NGC 6894 (PNG 069.4-02.6, Fig. \ref{lista2})} and {\bf NGC 7048
  (PNG 088.7-01.6, Fig. \ref{lista2})} are morphologically
simpler. The $\rm A_{v}$ of NGC 6894 is taken from Ciardullo et
al. (1999). The extinction coefficient determinations for NGC 7048 are
more contradictory.  By adopting the lowest one, we found that this is
a nearby nebula ($<$1000 pc). However, any distance closer than 1000
pc makes the central star absolute magnitude so faint that using a
standard evolutionary track gives an age of $>> 10^5$ yrs for the PN,
so it is likely more distant. Assuming instead the $c_{\beta}$ 
  obtained by our measurement, which agrees with the value in
Sabbadin et al. (1987) ($\rm A_v=1.27$), we found a distance between
1000 and 2000 pc.

{\bf NGC 7027  (PNG 084.9-03.4, Fig. \ref{lista2})} is the clearest case presenting evidence of strong internal
absorption, as discussed in section \ref{errori}. Given the difficulty
in estimating its exact amount as a function of the position at which
$c_{\beta}$ is measured, no extinction distance is given in
Tab. \ref{tav1} for this nebula.

{\bf NGC 7026 (PNG 089.0+00.3, Fig. \ref{lista2})} is a bipolar nebula with a low extinction coefficient. Using
any of the extinction values in Tab.~\ref{tav1}, 
we found a distance $<1500$ pc.

Finally, Benetti et al. (2003) propose {\bf NGC 7354 (PNG 107.8+02.3, Fig. \ref{lista2})} as a probable
recombining nebula in its first phase, as derived from a detailed
study of NGC 6818. Using the Feibelman (2000) extinction data  and extrapolating the extinction-distance curve to low distances,
we found a possible distance of about 1000 pc.

\section{Determination of distances for whole sample }

In this section distances are presented for 64 other planetary nebulae
in the area of sky covered by IPHAS. For a further 29 objects only a
lower limit to the distance can be obtained. For 17 PNe we found an
upper limit, as given in Table \ref{kstars}.  We adopt a distance
upper limit when the measured extinction to the nebula is smaller than
the first point on the corresponding distance-extinction curve. A
distance lower limit is defined when the visual extinction to the
nebula lies on the plateau of the interstellar extinction curve. For
multiple determinations of $c_{\beta}$ the source of the adopted value
is reported in bold-face. For upper and lower limits, if no indication
is given it means that the same conclusion holds for all tabulated
$c_{\beta}$'s.  The errors on the distances given in Table
\ref{kstars} are the fitting errors ${\sigma}_d$.

For the 14 PNe listed in Tab.~\ref{escluse}, we could not estimate the
distances because the measured extinction was too far above the
interstellar extinction plateau. According to the discussion of
section \ref{av} and the result presented in Fig. \ref{fig1}, we
assume that we know $ c_{\beta}$ with the uncertainty $\sigma$=0.41,
which is 0.88 mag in ${\rm A_v}$, for the data extracted from Acker et al. (1992).

So we consider all PNe with a visual extinction more than 0.88 magnitude
higher than the plateau value to live ``above" the plateau.  Considering these
nebulae above the plateau is a zero order hypothesis, because the measured
extinctions are not a homogeneous set of measurements and using a common sigma
is not rigorous. However, the PNe listed in Table \ref{escluse} form a sample
for future detailed studies, to understand whether the discrepancy with the
reddening-distance curve is due to measurement errors, or if the nebulae have
peculiar physical conditions that affect the determination of $c_{\beta}$
(e.g. very high densities which introduce optical thickness effects), whether
they are misclassified objects (symbiotic stars, nebulae around young
stars,etc.), or whether there is indeed a non negligible amount of reddening
associated with the PNe themselves and not of interstellar origin.

\begin{figure}
\centering
\resizebox{\hsize}{!}{\includegraphics{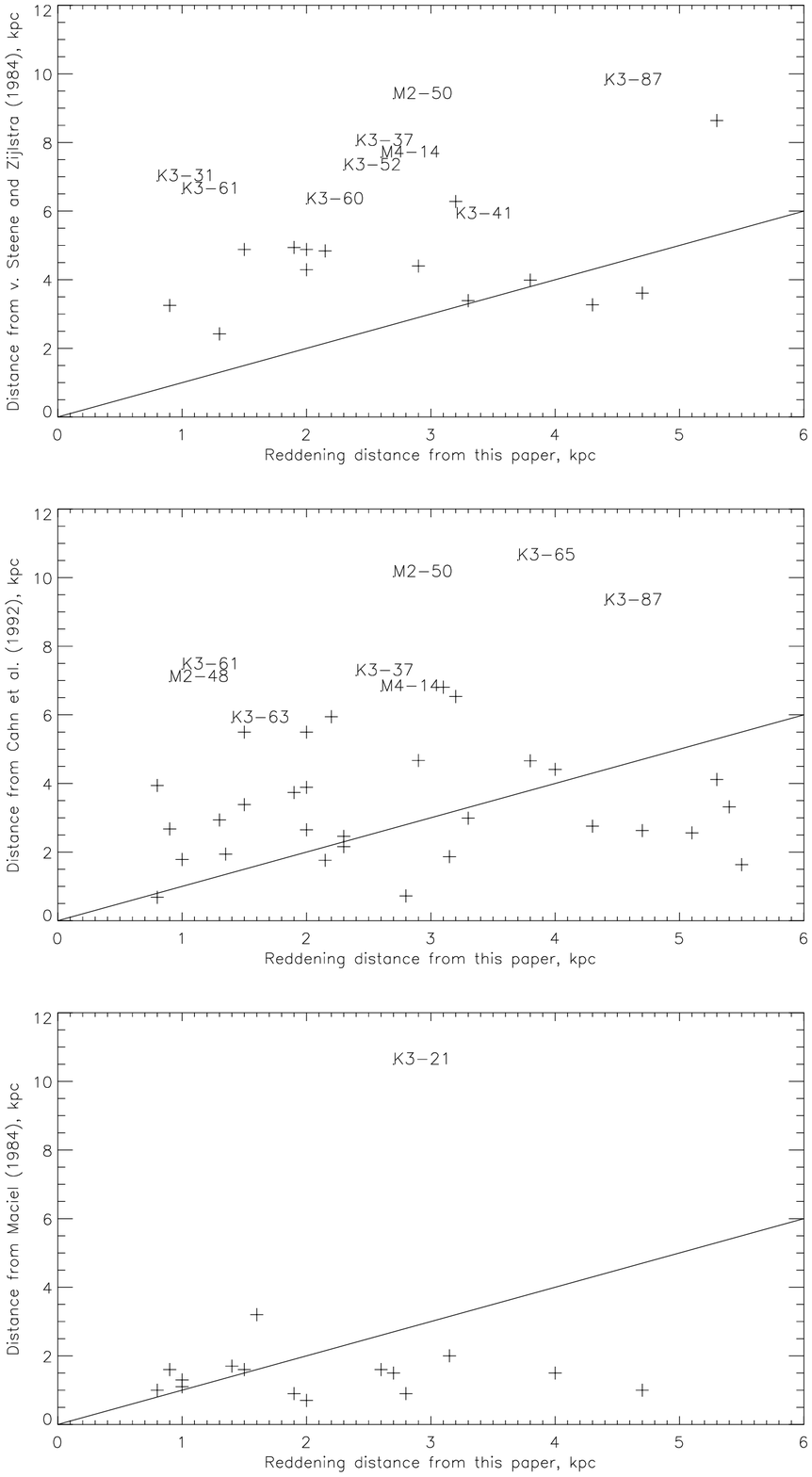}}
\caption{Comparison between the reddening
  distance and the statistical determination.The straight line is the
  1:1 locus, the statistical distance for the nebulae over the line is
  bigger than that inferred by the extinction.  Those PNe for which
  the difference between the two inferred distances is larger than 4
  kpc are labeled in the figure.}
\label{fig_d}
\end{figure}

The inferred distances are compared in Fig. \ref{fig_d} with those of
Cahn et al. (1992), based on a modified Shklovsky-Daub method, of
van de Steene \& Zijlstra (1994), based on a correlation between radio
continuum brightness temperature and radius, and of Maciel (1984),
  based on a mass--radius relationship established from selected
  electron densities and distances.  It is interesting to note that
the first two methods seem to provide larger distances than our
extinction method, while the latter seems to provide shorter
distances.  The spread in these graphs is 4.0~kpc (Cahn et
  al. 1992, 39 objects), 6.6~kpc (van de Steene \& Zijlstra 1994, 23
  objects), and 2.5~kpc (Maciel, 1984).  It therefore seems that the
  distances found by Maciel better agree with ours than those from the
  other authors. However, the small number of objects considered (16
  PNe, nine of which are at an estimated distance lower than 2 kpc), prevents 
any conclusion to be drawn with the present sample.


 Comparison with other distance scales would be very valuable to
  better understand the problem.  A detailed comparison with distance
  determined via the relation between size and $\rm H_\alpha$ surface
  brightness (Frew et al., 2006; Frew, 2008) will be presented in a future paper, when the
  global photometric calibration of the IPHAS survey becomes available.



\begin{table}[]
\caption{List of PNe with a reported visual extinction more than 1 mag
  larger than the plateau of the corresponding distance-extinction
  curve. The principal source for the $\rm H_{\alpha}$/ $\rm H_{\beta}$  ratio is the
  ESO/Strasbourg catalog marked as 1 in the column ``Reference'', while the
  other sources are: 
  2) Stasi\'nska et al. 1992,
  3) Cahn et al. 1992, 
  4) Manchado et al. 1989. $\dagger$ Note: K3-18 is a possible symbiotic star.}
\label{escluse}  \centering %
\begin{tabular} {c c c c c} 
PNG    &\multicolumn{1}{c}{Name}  &   $\rm A_v$ & \multicolumn{1}{c}{$\rm A_v$ Plateau} & \multicolumn{1}{c}{Reference}     \\
\hline \hline                 
032.0-03.0 & K3-18$\dagger$& 5.56 & 4.5 &1 \\ 
032.9-02.8 & K3-19 & 	4.49 & 3&1   \\
048.1+01.1 & K3-29 & 	6.31&  3.5&2\\ 
051.0+02.8 & WhMe1 & 	 7.60& 5  &1\\   
054.4-02.5 & M1-72 & 	 4.80& 2.5 &1, 3\\
055.2+02.8 & He2-432 &   6.32&  4 &1   \\
059.9+02.0 & K3-39 & 	  5.78&3.8 &1 \\
064.9-02.1 & K3-53     &    6.25&3.8 &1\\
069.2+02.8 & K3-49  & 4.76&3.4 &1\\
072.1+00.1 & K3-57 & 4.56&3 &1\\
094.5-00.8 & K3-83  & 4.72&3 &1\\
104.1+01.0 & Bl2-1  & 5.76& 4&1\\
110.1+01.9 & PM1-339  &  7.35 &6 &4 \\
112.5+03.7 & K3-88  & 5 & 4&1\\
 \hline
\end{tabular} 
\end{table} 

\section{Conclusions}

In this paper we have discussed the extinction-distance method for
determining distances to PNe, motivated by the opportunity provided by
the IPHAS survey with its wide application in the Galaxy. 

The technique is presented and the derivation of the distance and its
error is put on a formal mathematical basis.  We find a mean value of the
slope $b=6.53 \cdot 10^{-4}$ mag/pc for the galactic extinction
distance curve, valid for the analyzed region of the sky.

The numerical and physical reasons that can affect the measurement of
the interstellar extinction toward a PN are then discussed. All these,
which add to the unavoidable measurement errors the effects of an
incomplete knowledge of the physics of interstellar dust through the
Galaxy and inside PNe, have to be taken into account every time the
method is applied.


In spite of these uncertainties which will need to be addressed in
future, we have computed extinction distances for a sample of 70 PNe,
which is the largest sample of Galactic PNe to which the technique has
been applied.  In the best conditions, when the extinction
measurements of the PNe are of good quality and the lines of sight are
well behaved in terms of their associated distance-extinction maps,
the errors in the PN distances can be as good as 20\%, which is a very
promising result for a wider application of the method.  Unfortunately
this is not the common situation for the current data set. As
previously mentioned, the uncertainty on the visual extinction for the
whole set of nebulae is $\sigma=0.88$, which implies much larger
errors in the distances. For a subsample of the PNe considered,
  extinction is more precisely measured, and for the two cases for which
  we could determine the associated error, an uncertainty of 35\%\ on
  the distance is estimated. This can be taken as a figure to be
  associated with PNe for which good spectroscopic data exist. 
A main message of
this study is that a careful determination of the nebular extinction
is mandatory to limit the errors in a way that reliable distances can
be obtained for individual objects. In this respect, our plan is to
improve the extinction measurements for a significant sample of
Galactic PNe.

We have then compared these new extinction distances with other
statistical methods. Some of the methods considered (Cahn et al. 1992,
and van de Steene \& Zijlstra 1994) provide distances that are
generally larger than our determinations, and in general the
dispersion is quite high. This emphasizes the large 1-$\sigma$ dispersion in 
the statistical distances, and the dangers in relying on
the determination of a statistical distance to a PN without considering the
particular characteristic of a nebula.


Concluding, the analysis in this paper demonstrates the increased
potential of the extinction method gained with the availability of
surveys like IPHAS, which provide precise photometric data in large
areas of sky. This opens the possibility of making a significant step
forward in the calibration of the PN distance scale in the Galaxy, and
in our understanding of important issues such as the properties of the
dust distribution within the Galactic disc, and the possible
importance of dust associated to PNe.

\begin{acknowledgements}

CG, RLMC, AM, and MSG acknowledge funding from the Spanish AYA2007-66804 grant.  

\end{acknowledgements}

\clearpage
\onecolumn
\begin{center}
\begin{scriptsize}
\label{tavola}
\begin{longtable}{c|c|c c c c c| c c c c|c|c}  
\caption{\label{kstars} Distance to the whole sample of PNe (except
  the NGC objects listed in the Table \ref{tav1} and the objects listed in Table \ref{escluse}). The columns numbered  from 1 to 3 give
  the values of $c_{\beta}$ inferred from  H$_{\alpha}$ and H$_{\beta}$ supplied to the Strasbourg
  catalog (1),  from Cahn et al. (1992) (2), from Stasi\'nska et
  al. (1992) (3). The extinctions obtained in this paper are given in column 4. In column 5 we present the values obtained in other papers, the corresponding references are listed at the end of the table, and the corresponding sequential  numbers are shown in column  ``key''.
  The values adopted in order to obtain the distances are marked in bold-face. If for a PNe no value is shown in bold-face, it means that we find the same results with all of them.
 Our distance are compared with  those obtained by Cahn et al. (1992), van de Steene \& Zijlstra  (1994) and Maciel (1984).}\\

     PNG & Name &   \multicolumn{5}{c}{$c_\beta$} &  \multicolumn{4}{c}{distance [pc] } & key\\  
              &	    &  Stras. & Cahn &  Stas.   &this p.& Other &  Cahn   & Vst. & Ma. &  THIS   &                  \\
              &	    &  (1)    & (2)  &  (3)     & (4)   & (5)   &         &      &     &  PAPER  &                  \\
\hline
\hline
\endfirsthead
\caption[]{Continued} \\
     PNG    &	       Name  &Stras.  & Cahn &  Stas.   &this p.& Other &  Cahn   & Vst.  & Ma. &  this p.&  key   \\
              &	    &  (1)    & (2)  &  (3)     & (4)   & (5)   &         &      &     &  PAPER  &                  \\
\hline
\hline
\endhead 
\hline
\multicolumn{13}{c}{continued on next page} \\
\endfoot
\hline
\endlastfoot
\hline
   031.7+01.7 &          PC20 &    2.51&        &        &    3.04&        &        &        &        &  $>$1100                &\\    
  032.5-03.2 &         K3-20 &    1.88&        &        &        &        &        &        &        &  $>$1100                &\\    
  032.7-02.0 &         M1-66 &    1.39&        &        &    1.44&        &   3439 &   5000 &        &  $>$3000                &\\    
  033.2-01.9 &       Sa3-151 &        &        &        &    1.52&        &        &        &        &   3500  $\pm$ 300       &\\    
  035.9-01.1 &        Sh2-71 &    1.65&    1.38&        &    1.18&        &    997 &        &        &  $<$1000                &\\    
  038.7-03.3 &         M1-69 &    1.32&        &        &        &        &        &        &        &   3300  $\pm$ 500       &\\    
  040.4-03.1 &         K3-30 &    1.67&        &    1.66&        &        &   6291 &   6550 &        &  $>$3000                &\\    
  041.2-00.6 &        HaTr14 &    0.56&        &        &        &        &        &        &        &  $<$1000                &\\    
  041.8+04.4 &         K3-15 &    1.28&        &        &        &        &        &        &        &  $>$6000                &\\    
  043.0-03.0 &         M4-14 &    1.42&        &        &        &        &   6692 &   7570 &   1600 &   2600  $\pm$ 300       &\\    
  043.1+03.8 &         M1-65 &    1.24&        &  \bf  1.23&        &        &   6536 &   6280 &        &   3200  $\pm$ 900       &\\    
  043.3+02.2 &       PM1-276 &        &        &        &        &    1.74&        &        &        &   1350  $\pm$ 100       &1\\   
  045.4-02.7 &         Vy2-2 &    1.94&   \bf 1.71&        &        &        &   2159 &        &        &   2300  $\pm$ 170       &\\    
  046.3-03.1 &           PB9 &    1.81&        &        &        &        &   4661 &   3990 &        &   3800  $\pm$ 450       &\\    
  047.1+04.1 &         K3-21 &    1.15&        &        &        &        &        &        &  10500 &   2700  $\pm$ 366       &\\    
  047.1-04.2 &           A62 &        &    2.07&        &        & \bf   0.18&    494 &        &   1100 &  $<$900   	       &2\\         
  048.0-02.3 &          PB10 &    2.01&        &        &        &        &   4685 &   3830 &        &  $>$2000 	       &\\           
  048.5+04.2 &         K4-16 &    1.51&        & \bf   1.50&        &        &  15127 &        &        &  $>$5000 	       &\\           
  048.7+01.9 &       He2-429 &    2.12&        &        &        & \bf   1.87&   3383 &        &        &   1500  $\pm$ 600       &3\\   
  049.4+02.4 &       He2-428 &    1.23&        &        &        &   \bf 1.14&        &        &        &   1000  $\pm$ 100       &7\\   
  050.1+03.3 &         M1-67 &    1.07&        &        &        &        &    682 &        &   1000 &   800   $\pm$ 100       &\\    
  051.0+03.0 &       He2-430 &    2.22&        &    2.21&        &        &   3972 &        &        &  $>$3000 	       &\\           
  051.0-04.5 &          PC22 &    0.58&        &        &        &        &        &        &        &   4000  $\pm$ 500       &\\    
  051.9-03.8 &         M1-73 & \bf   0.88&        &        &        &    0.95&   4669 &   4400 &        &   2900  $\pm$ 500       &4\\   
  052.2-04.0 &         M1-74 &    0.94&    1.01&   \bf 0.98&        &        &   4118 &   8640 &        &   5300  $\pm$ 1000      &\\    
  052.5-02.9 &         Me1-1 &    0.05&    0.72&    0.46&        &  \bf  0.58&   4618 &        &        &  $<$1000 	       &5\\          
  052.9+02.7 &         K3-31 &    2.32&        &        &        &        &   3941 &   6880 &        &   800   $\pm$ 700       &\\    
  052.9-02.7 &         K3-41 &    1.13&        &        &        &        &  20464 &  28690 &        &   3200  $\pm$ 500       &\\    
  053.3+03.0 &           A59 &        &    1.77&        &    1.61&        &   1412 &        &   1800 &  $<$1000 	       &\\           
  053.8-03.0 &           A63 &    1.12&        &        &    1.09&        &        &        &   2700 &  $>$8000  	       &\\          
  055.1-01.8 &         K3-43 &    1.66&        &        &        &        &        &        &   1500 &  $>$2800 	       &\\           
  055.3+02.7 &         He1-1 &    2.06&        &        &        &        &        &        &        &   2500  $\pm$ 1000      &\\    
  055.5-00.5 &         M1-71 &    1.82&        &    1.80&        &  \bf  2.06&        &        &        &   2900  $\pm$ 400       &6\\   
  055.6+02.1 &         He1-2 &    1.86&        &        &        &        &        &        &        &  $>$3000  	       &\\          
  056.0+02.0 &         K3-35 &    2.01&        &        &        &        &   3975 &   6580 &        &  $<$1000   	       &\\         
  057.9-01.5 &       He2-447 &    2.18&        &        &        &        &   2874 &        &        &  $>$2000   	       &\\         
  058.9+01.3 &         K3-40 &    1.72&        &        &        &        &   7068 &   6490 &        &  $>$3000   	       &\\         
  059.0+04.6 &         K3-34 &    0.35&        &        &        &        &        &        &   5700 &  $<$1000   	       &\\         
  059.0-01.7 &         He1-3 &    1.67&        &        &        &        &        &        &        &   1000  $\pm$ 200       &\\    
  059.4+02.3 &         K3-37 &    1.77&        &        &        &        &   7148 &   7910 &        &   2400  $\pm$ 400       &\\    
  060.5-00.3 &         K3-45 &    0.84&        &        &        &        &        &        &   1900 &  $<$1000  	       &\\          
  060.5+01.8 &       He2-440 &    2.94&        &        &        &        &   3962 &        &        &  $>$4000     	       &\\       
  061.3+03.6 &       He2-437 &    1.45&        &        &        &   \bf 0.88&        &        &        &   1100  $\pm$ 300       &7\\   
  062.4-00.2 &         M2-48 &    2.02&        &        &        &   \bf 1.35&   6970 &        &   1600 &   900   $\pm$ 200       &8\\   
  067.9-00.2 &         K3-52 &    1.49&        &        &        &        &   2459 &   7200 &        &   2300  $\pm$ 200       &\\    
  068.6+01.1 &         He1-4 &    1.77&        &        &    1.44&        &        &        &        &  $>$5000  	       &\\          
  068.7+01.9 &         K4-41 &    1.50&        &        &        &        &   7929 &   7820 &        &  $>$4000  	       &\\          
  068.7+03.0 &          PC23 &    1.53&        &        &        &        &   6680 &   7740 &        &  $>$4000   	       &\\         
  068.8-00.0 &         M1-75 &        &        &        &    2.11&        &   3190 &        &   3100 &  $>$4000 	       &\\           
  069.2+03.8 &         K3-46 &    0.69&        &        &        &        &        &        &   3200 &   1600  $\pm$ 100       &\\    
  069.6-03.9 &         K3-58 &    1.34&        &        &        &        &   5635 &        &   5700 &  $>$3000  	       &\\          
  071.6-02.3 &         M3-35 &   \bf 2.07&    2.41&        &        &        &   1760 &   4840 &        &   2150  $\pm$ 150       &\\    
  075.6+04.3 &    Anon.20h02 &    0.01&        &        & \bf   0.85&        &        &        &        &   7000  $\pm$ 1300      &\\    
  076.3+01.1 &          A69  &        &  4.35     &        &        &        &   4173 &        &    300 &  $>$4000  	       &\\          
  076.4+01.8 &         KjPn3 &    0.93&        &        &        &        &        &        &        &   2000  $\pm$ 100       &\\    
  077.5+03.7 &         KjPn1 &    1.54&        &        &        &        &        &        &        &  $>$5500  	       &\\          
  077.7+03.1 &         KjPn2 &    1.52&        &        &        &        &        &        &        &   3700  $\pm$ 400       &\\    
  078.9+00.7 &           Sd1 &    0.11&        &        &        &    0.11&        &        &        &  $<$1200  	       &9\\         
  084.2-04.2 &         K3-80 &    1.18&        &        &        &        &        &        &        &   900   $\pm$ 100       &\\    
  084.9+04.4 &           A71 &        &    1.06&        &        &        &    722 &        &    900 &   2800  $\pm$ 400       &\\    
  088.7+04.6 &         K3-78 &    0.35&        &        &        &        &   7831 &   7290 &        &  $<$1200  	       &\\          
  089.3-02.2 &         M1-77 &    1.23&        &        &        &        &   5496 &   4880 &    700 &   2000  $\pm$ 250       &\\    
  089.8-00.6 &        Sh1-89 &    0.99&    \bf 0.82&        &        &        &   1941 &        &        &   1350  $\pm$ 200       &\\    
  091.6-04.8 &         K3-84 &    0.51&        &        &        &        &        &        &        &   5200  $\pm$ 900       &\\    
  093.3-00.9 &         K3-82 &    0.96&        &        &        &        &   2675 &   3250 &        &   900   $\pm$ 130        &\\   
  093.3-02.4 &         M1-79 &    0.13&    0.78&  \bf  0.77&        &        &   2627 &   3610 &   1000 &   4700  $\pm$ 800        &\\   
  093.5+01.4 &         M1-78 &    3.02&    2.65&        &        &        &    700 &        &   3100 &  $>$2200 	        &\\          
  095.1-02.0 &         M2-49 &    1.17&        &        &        &        &   2870 &   6140 &        &  $>$2300 	        &\\          
  095.2+00.7 &         K3-62 &    2.54&        &        &        &        &   2270 &   4050 &        &  $>$2500 	        &\\          
  096.3+02.3 &         K3-61 &    1.32&        &        &        &        &   7334 &   6510 &   1100 &   1000  $\pm$ 200        &\\   
  097.6-02.4 &         M2-50 &    0.95&    0.63&   \bf 0.61&        &        &  10028 &   9280 &   1500 &   2700  $\pm$ 500        &\\   
  098.1+02.4 &         K3-63 &    1.18&  \bf  0.99&        &        &        &   5780 &        &   1700 &   1400  $\pm$ 130        &\\   
  098.2+04.9 &         K3-60 &    2.27&    1.90&    2.59&        &        &   3887 &   6210 &        &   2000  $\pm$ 200        &\\   
  102.9-02.3 &           A79 &        &    1.33&        &    0.81&  \bf  0.36&   1784 &        &   1300 &   1000  $\pm$ 160        &7\\  
  103.7+00.4 &         M2-52 &    1.72&        &        &    1.44&        &   4411 &        &   1500 &   4000  $\pm$ 360        &\\   
  104.4-01.6 &         M2-53 &    0.52&    1.26&        &    1.06&   \bf 0.73&   3737 &   4940 &    900 &   1900  $\pm$ 400        &\\   
  107.4-00.6 &         K4-57 &    1.32&        &        &        &        &        &        &        &   4000  $\pm$ 1000       &\\   
  107.4-02.6 &         K3-87$\dagger$ &    1.14&        &        &        &        &   9203 &   9680 &        &   4400  $\pm$ 800        &\\   
  107.7-02.2 &         M1-80 &    0.52&        &        &        &        &   5496 &   4880 &   1600 &   1500  $\pm$ 200        &\\   
  111.8-02.8 &          Hb12 &        &        &    1.35&        &    0.60&        &        &        &  $<$1000  	        &11\\       
  112.5-00.1 &         KjPn8 &    0.67&        &        &        &  \bf  0.42&        &        &        &   1400  $\pm$ 100        &17\\ 
  114.0-04.6 &           A82 &        &    0.75&        &        &        &   1868 &        &   2000 &   3150  $\pm$ 500        &\\   
  119.3+00.3 &         BV5-1 &    1.05&        &    1.31&    1.01&\bf    0.82&        &        &        &   3000  $\pm$ 400        &12\\ 
  122.1-04.9 &            A2 &    0.76&    \bf 0.56&    0.34&    0.80&        &   3929 &        &   3000 &  1500 $\pm$ 150 	 &\\          
  126.3+02.9 &         K3-90 &    0.83&    0.97&        &        &        &   5759 &   5600 &        &  $<$1000  	        &\\         
  128.0-04.1 &      Simeiz22 &        &        &        &        &    0.34&        &        &        &  $<$1000   	        &13\\      
  130.2+01.3 &        IC1747 &    1.48&    0.74&        &    1.03&        &   2937 &   2420 &        &   1300  $\pm$ 200        &\\   
  130.4+03.1 &         K3-92 &    1.96&    1.30&        &  \bf  1.23&        &   6806 &        &        &   3100  $\pm$ 900        &\\   
  131.5+02.6 &            A3 &        &    1.23&    1.22&    1.33&   \bf 1.23&   2560 &        &        &   5100  $\pm$ 500        &14\\ 
  132.4+04.7 &         K3-93 &  \bf  1.35&        &        &    1.94&        &        &        &        &  $<$1000 	        &\\          
  136.1+04.9 &            A6 &        &    1.39&        &        &    1.39&    958 &   2240 &        &  $<$1000  	        &\\         
  138.8+02.8 &         IC289 &    1.03&    1.09&        &    1.16&        &   1434 &   1480 &        &  $>$8000   	        &\\        
  142.1+03.4 &         K3-94 &    0.91&        &        &        &  \bf  0.52&   6507 &   7740 &        &  $<$1000 	        &15\\        
  147.4-02.3 &          M1-4 &    2.47&    1.84&   \bf 1.62&        &        &   2988 &   3390 &        &   3300  $\pm$ 350        &\\   
  147.8+04.1 &          M2-2 &    1.34&    1.50&    1.50&        &        &   4356 &   3880 &        &  $>$2000  	        &\\         
  151.4+00.5 &         K3-64 &    0.73&        &        &        &        &        &        &        &   1800  $\pm$ 200        &\\   
  153.7-01.4 &         K3-65 &    1.83&        &        &        &        &  10512 &        &        &   3700  $\pm$ 300        &\\   
  160.5-00.5 &         We1-2 &        &        &        &        &    1.44&        &        &        &   2500  $\pm$ 180        &16\\ 
  163.1-00.8 &         We1-3 &        &        &        &        &    0.70&        &        &        &   2700  $\pm$ 300        &16\\ 
  167.0-00.9 &            A8 &        &    1.77&        &        &  \bf  0.68&   1633 &        &        &   5500  $\pm$ 1000       &2\\  
  170.7+04.6 &         K3-69 &    1.34&        &        &        &        &   7946 &        &  17850 &  $>$6000 	        &\\          
  178.3-02.5 &         K3-68 &    0.80&        &        &        &        &   5945 &        &        &   2200  $\pm$ 240        &16\\ 
  181.5+00.9 &           Pu1 &        &        &        &        &    1.03&        &        &        &   7000  $\pm$ 800        &\\   
  184.0-02.1 &          M1-5 &        &    1.39&    1.38&        &        &   2922 &   4940 &        &  $>$4000   	        &\\        
  184.6+00.6 &         K3-70 &    1.55&        &        &        &        &  12117 &        &        &  $>$6000  	        &\\         
  184.8+04.4 &         K3-71 &    1.14&        &        &        &        &        &        &        &   2500  $\pm$ 1000       &\\   
  194.2+02.5 &          J900 & \bf   0.56&    0.64&    0.63&        &        &   2756 &   3270 &        &   4300  $\pm$ 650        &\\   
  197.8-03.3 &           A14 &        &    1.18&        &    1.52&  \bf  0.85&   3317 &        &        &   5400  $\pm$ 800        &10\\ 
  201.9-04.6 &         We1-4 &        &        &        &        &    0.85&        &        &        &   3000  $\pm$ 1100       &15\\ 
  204.8-03.5 &         K3-72 &        &        &        &        &    1.06&   5119 &        &        &  $>$4000  	        &15\\       
  211.2-03.5 &          M1-6 &    1.84& \bf   1.17&    1.83&        &        &   2648 &   4290 &        &   2000  $\pm$ 160        &\\   
  \end{longtable}
{key: 1) van de Steene et al. 1996;
2) Phillips et al. 2005;
3) Girard et al. 2007;
4) Wesson  et al. 2005;
5) Shen et al. 2004;
6) Wright et al. 2005;
7) Rodrigu\'ez et al. 2001; 
8) L\'opez-Mart\'{\i}n  et al. 2002;
9) Kazarian et al. 1998;
10) Bohigas  2003;
11) Rudy et al. 1993;
12) Bohigas 2008;
13) Kwitter \& Jacoby 1989;
14) Kaler 1983;
15) Bohigas 2001;
16) Kaler et al. 1990;
17) Gon\c calves et al. 2009.\\
$\dagger$ Note: K3-87 is a possible symbiotic.}
\end{scriptsize}
\end{center}

\end{document}